\documentclass[runningheads]{llncs}

\usepackage[T1]{fontenc}
\usepackage{graphicx}
\usepackage[utf8]{inputenc}
\begin{document}
\setlength{\belowcaptionskip}{-15pt}

\title{What Makes an Educational Robot Game Fun? Framework Analysis of Children's Design Ideas}
\titlerunning{What Makes an Educational Robot Game Fun?}
\author{Elaheh Sanoubari\inst{1,2}
\and
John Edison Mu\~{n}oz\inst{2}
\and
Ali Yamini\inst{1}
\and
Neil Randall\inst{2}
\and
Kerstin Dautenhahn\inst{1}}
\authorrunning{E. Sanoubari et al.}

\institute{Faculty of Engineering, University of Waterloo, Canada \and Games Institute, University of Waterloo, Canada \\
\email{\{esanouba, jmunozca, ayamini, nrandall, kdautenh\}@uwaterloo.ca}}
\maketitle          
\newcommand{\REMind}{\textit{REMind}}
\begin{abstract}
Fun acts as a catalyst for learning by enhancing motivation, active engagement and knowledge retention. As social robots gain traction as educational tools, understanding how their unique affordances can be leveraged to cultivate fun becomes crucial. This research investigates the concept of fun in educational games involving social robots to support the design of \REMind{}:
\footnote{\REMind{} is short for \textbf{R}obots \textbf{E}mpowering \textbf{M}inds}
a robot-mediated role-play game aimed at encouraging bystander intervention against peer bullying among children. 
To incorporate fun elements into design of \REMind, we conducted a user-centered Research through Design (RtD) study with focus groups of children to gain a deeper understanding of their perceptions of fun. We analyzed children's ideas by using Framework Analysis and leveraging LeBlanc's Taxonomy of Game Pleasures and identified 28 elements of fun that can be incorporated into robot-mediated games.
We present our observations, discuss their impact on \REMind's design, and offer recommendations for designing fun educational games using social robots.

\end{abstract}
\section{Introduction}
\vspace{-0.2cm}
Social robots are increasingly popular in education, offering unique opportunities to enhance learning. Their physical embodiment and expressive capabilities make them highly engaging, allowing for creating interactive and emotionally resonant experiences. 
By integrating social robots with educational games, we can create playful scenarios that offer personalized and situated experiences, increasing learner engagement and potentially enhancing learning outcomes \cite{rato2023robots}.

Games, as structured play, are excellent facilitators of learning because they induce Flow state~\cite{plass2020handbook,csikszentmihalyi2014toward}. Flow states, characterized by deep involvement with an activity and a distorted sense of time, are important in educational settings as they reduce self-consciousness and enhance learning \cite{plass2015foundations}. Furthermore, games create a safe space for learning by eliminating real-life costs of failure, encouraging persistence and self-regulated learning through multiple attempts and immediate feedback \cite{schell2008art}. Additionally, games engage learners as active producers of knowledge, unlike the passive consumption often found in traditional education \cite{gee2003video}. Thus, we argue that integrating Game-Based Learning (GBL) into social robot educational experiences could increase their effectiveness.

Broadly defined, GBL involves the use of games and game-like elements to support the acquisition of knowledge, skills, and attitudes. It differs from gamification, which typically involves adding incentives like stars, points, or rankings to motivate learners to engage with otherwise tedious tasks \cite{plass2020handbook}. 
Superficially incorporating fun elements into educational components (e.g., by merely transforming a quiz into a digital format with a point system) often falls flat and leads to disengagement if the core task is not fun. Game scholars use the metaphor of \textit{`chocolate-covered broccoli'} to illustrate this point: adding fun to a game as an afterthought is like coating vegetables in chocolate; it does not make them candy~\cite{supriana2017serious}. In contrast, GBL involves redesigning learning tasks to be inherently interesting and meaningful, and aiming to create experiences that are both effective educational tools and fun games \cite{plass2020handbook,kapp2012gamification}. 

While educational video games are a modern example, GBL has a long history. Developmental psychology has long recognized play as a natural form of learning, with significant research on playful learning predating the digital era \cite{plass2015foundations}. A key feature of play is that it is intrinsically motivating-- we play because it is fun. Games are engaging and enjoyable by design, which can increase learners' motivation and interest in the learning content, leading to higher levels of effort, persistence, and achievement \cite{kebritchi2010effects,malone1981makes}. 
Although games that have a purpose beyond entertainment are commonly labeled `Serious Games,' game design scholars argue that this label is misleading, as it implies that such games need not prioritize fun; instead they advocate for the term `Transformational Games'~\cite{schell2008art}. Regardless of terminology, to be effective, such games must indeed be fun \cite{plass2020handbook}. 
Hence, it is crucial to explore how we might objectively integrate fun into the designs of educational experiences. Game designers highlight several key factors: goals, feedback, interactivity, storytelling, aesthetics, challenges, and social interaction. While the fundamentals of play remain consistent, we argue that novel technologies like social robots introduce new opportunities for play through their unique affordances. This raises the question: \textbf{how can social robots make transformational games more fun?}

We explored this question in the context of developing \REMind, an educational game using social robots to teach anti-bullying intervention skills to children. The subjective and context-dependant nature of fun poses a design challenge. Recognizing users as ``experts of their own experiences''~\cite{sanders2008co}, we took a user-centered Research through Design (RtD) approach. We engaged children, the game's primary stakeholders, in focus groups to brainstorm ways to make \REMind{} fun.
We analyzed the qualitative data gathered from this study by applying Framework Analysis~\cite{goldsmith2021using} and using LeBlanc's Taxonomy of Game Pleasures~\cite{hunicke2004mda}, which identifies 8 categories of enjoyment in games. This method helped us identify 28 elements of fun in robot games, which informed our game design. In this paper, we present our findings, discuss how they shaped the design of \REMind, and offer design recommendations based on these insights.
\section{Related Work}
\vspace{-0.3cm}
Social robots are emerging as a promising tool with the potential to enhance learning experiences in both classrooms and homes~\cite{belpaeme2021social}. A growing body of research shows that social robots can enhance children's cognitive, conceptual, language, and social development~\cite{toh2016review}. Studies have demonstrated improvements in subjects such as math and science, as well as skills like teamwork and social interaction when robots are integrated into educational environments~\cite{kandlhofer2016evaluating}. Specifically, robots have been found to boost motivation, attention, and creativity among learners, positively impacting their overall educational experience \cite{belpaeme2018social,kahn2016human}. 
Additionally, studies have shown that incorporating robots into mental health interventions can reduce distress and increase positive affect in children~\cite{kabacinska2021socially}.

Social robots possess unique attributes that might make them well-suited for educational purposes. Their social presence and life-like behavior can elicit pro-social behavior and enhance engagement among learners~\cite{kim2013social}. This is especially beneficial for children on the autism spectrum, where robots' social responsiveness and non-judgmental nature can reduce anxiety and foster a more conducive learning environment~\cite{dautenhahn2007socially}. Robots' infinite patience and ability to provide personalized content and feedback further enhance their effectiveness \cite{belpaeme2021social}. Importantly, research shows that robots can elicit playful behavior in children \cite{dautenhahn2007socially}. This is significant as play is recognized as the ``most natural form of learning'' and seminal psychologists such as Piaget and Vygotsky have emphasized the key role of play in children's development~\cite{plass2020handbook}.
Structuring educational experiences with social robots as games leverages their unique attributes to create engaging learning.

Fun is often cited as the primary reason for integrating games into learning environments~\cite{plass2020handbook}, as it supplies learners with intrinsic motivation. Fun is defined as an enjoyable experience that extends beyond a jolt of pleasure, which individuals typically seek to repeat~\cite{sellers2017advanced}. Schaufeli et al., describe fun as a ``pervasive affective-cognitive state,''~\cite{schaufeli2002measurement} while Koster, in his influential book ``Theory of Fun,'' asserts that ``fun is just another word for learning''~\cite{koster2013theory}. Research has consistently shown that positive emotions associated with fun not only broaden cognitive resources \cite{fredrickson2005positive} but also enhance learners' attentive state \cite{izard1993four}, serve as effective retrieval cues \cite{isen1978affect,isen1987positive}, and improve decision-making, creative problem-solving, and other higher-level cognitive functions \cite{erez2002influence}. 

Incorporating fun into the design of a system is challenging because fun is an inherently subjective experience. 
Related work on video game motivation identified challenge, fantasy, and curiosity as key components of fun \cite{malone1981makes}.
Lazzaro, by analyzing players' facial expressions, identified four types of fun: Hard Fun (emphasizing challenge and mastery), Easy Fun (exploration and creativity), Serious Fun (aimed at learning and personal growth), and People Fun (social interactions and community building) \cite{lazzaro2009we}. Furthermore, game designer Marc LeBlanc has outlined eight primary ``game pleasures'' that encapsulate the main sources of enjoyment in games \cite{hunicke2004mda}: Sensation, Fantasy, Narrative, Challenge, Fellowship, Discovery, Expression, and Submission (see Table \ref{tab:8fun_definition} for definitions). 
Others have expanded this list to as many as 14 types of fun\cite{garneau2001fourteen}.
These frameworks illustrate the multi-faceted nature of fun and provide design insights.

\section{Designing REMind}
\vspace{-0.2cm}
\label{intro_to_REMind}
This research supports the ongoing development of \REMind, a transformational game that aims to use role-playing with social robots to encourage anti-bullying bystander intervention in children. \REMind{} is structured as a room-based installation in which a single player interacts with 3 Furhat robots. Furhat is a tabletop human-like social robot with a back-projected head capable of displaying a range of facial expression~\cite{al2012furhat}. The choice of Furhat as the platform for \REMind{} was informed by findings from a prior co-design research \cite{sanoubari2021robots}.

\REMind{} engages players in an interactive story where the mission is to assist a robot named ECHO in bystander intervention. This concept is inspired by Forum Theatre~\cite{boal2002games}, a theatrical game where spectators of a drama are invited to become "spect-actors." In this format, the audience watches the performance twice; during the second viewing, they can stop the performance and suggest different actions for the protagonist.
\REMind{} uses spect-actorship as a game mechanic, allowing children to first observe a bullying scenario between two robots, then guide ECHO in intervening as the bystander.

Research shows that learning is most effective when game mechanics are clearly aligned with learning objectives~\cite{plass2020handbook}. 
Based on previous interviews with teachers~\cite{sanoubari2022designing} and consultations with expert psychologists, we identified the following learning goals for 
\REMind: (a) help players differentiate between bullying and teasing by understanding that bullying has 3 criteria \cite{cuadrado2012repetition} (attributed to Olweus): imbalance of power, intention to harm, and repetition; (b) foster empathy and perspective-taking; and (c) promote safe and effective intervention strategies such as assertive defending, comforting victims, or reporting to authorities.
To enhance the effectiveness of \REMind, we set out to understand how its game mechanics can be further developed in alignment with these learning goals while ensuring fun gameplay.

\section{Methodology}
\vspace{-0.2cm}
We took a user-centered Research through Design (RtD) approach. RtD is a methodology from Human-Computer Interaction (HCI) that uses the design process to explore research questions and generate tacit understanding~\cite{gaver2012should}. 
To directly involve children in the design process, we engaged them in focus groups in which we introduced them to \REMind{} and its educational objectives, demonstrated the robots' capabilities and engaged in a group brainstorming activity.

\vspace{-0.3cm}
\subsection{Procedure}
Two researchers facilitated the design sessions. After reaffirming children's assent and a brief ice-breaker activity, we explained \REMind{}'s goal of fostering anti-bullying intervention and its core mechanic of role-playing with robots. To spark children's imagination, we provided detailed game character sheets (compiled in collaboration with a professional playwright based on findings in a previous design study~\cite{sanoubari2021robots}). To ground children's design ideas in the affordances of Furhat robot, we presented them with a brief demo, showcasing robot's ability to communicate, change its appearance, and manage information. 

For the brainstorming activity, we used the \textit{Crazy Eights}~\cite{crazy8s} technique for rapid idea generation. In this activity, each child was tasked to reflect on `How might we make \REMind{} fun' and create eight distinct ideas in eight minutes, scribbling each on a sticky note. Afterwards, children shared their ideas with the group. Each idea was placed on a large whiteboard, and peers could indicate if they had similar thoughts as we visually clustered them based on themes. Researchers posed follow-up questions when necessary.

Each design session lasted one hour. As a token of appreciation for their contribution to \REMind{}, children received personalized certificates and their parents received CAD \$15. This protocol was approved by the the University of Waterloo's Human Research Ethics Board. 

\vspace{-0.3cm}

\subsection{Recruitment \& Participants}
\vspace{-0.2cm}
We promoted the study through local mailing lists and interested parents contacted us and received a detailed information letter. Parents of 15 children aged 8-14 (M=11, SD=1.64) provided consent and confirmed child assent.
Following recommendations on optimal focus group size \cite{stewart2007focus}, participants were divided into two groups based on their availability. One group consisted of 7 children aged 8-11, and the other group included 8 children aged 10-14.
We did not collect information about gender, but the groups seemed roughly gender-balanced.

\vspace{-0.3cm}

\subsection{Data \& Analysis}
\vspace{-0.2cm}
Data from this study included 95 game design ideas on sticky notes generated by  the children (M=6.33 ideas per child, SD = 3.06) and 2 hours of video recordings.

We used Framework Analysis~\cite{goldsmith2021using}, a systematic methodology in qualitative research, to interpret this data. First, we thoroughly reviewed the data and created an annotated dataset. Then, we identified LeBlanc's Taxonomy of Game Pleasures \cite{hunicke2004mda} to develop a coding scheme (see Table \ref{tab:8fun_definition}). Next, we deductively coded each game design idea into relevant game pleasure categories. Finally, we identified sub-themes within each category and analyzed their key features. The first author conducted this analysis was conducted using NVIVO \cite{nvivoR1}. 
To evaluate Intercoder Reliability, the second author  deductively coded a 25\% subset of the dataset, as recommended by literature~\cite{o2020intercoder}. This resulted in an agreement of 0.72 and a Cohen's Kappa of 0.66, indicating substantial agreement.

\begin{table}[h!]
\centering
\caption{LeBlanc proposes a taxonomy of eight primary game pleasures~\cite{hunicke2004mda,schell2008art}.}\label{tab:8fun_definition}
\resizebox{0.95\columnwidth}{!}{
\begin{tabular}{ll}
\hline
\textbf{Pleasures} & \textbf{Definition}                                                                     \\ 
\hline
Sensation          & Enjoyment from sensory experiences (visuals, music, tactile feedback).       \\ 
\hline
Fantasy            & Pleasure derived from imagining oneself in different roles or worlds.                   \\ 
\hline
Narrative          & Enjoyment of unfolding events or stories.          \\ 
\hline
Challenge          & The core pleasure of solving problems within a game.                                    \\ 
\hline
Fellowship         & Enjoyment from friendship, cooperation, and community.                  \\ 
\hline
Discovery          & Pleasure from finding new things and exploring game worlds.        \\ 
\hline
Expression         & Joy of self-expression \& creativity, e.g., designing characters or levels.  \\ 
\hline
Submission         & Pleasure from immersing oneself in the game world and escaping reality.                 \\
\hline
\end{tabular}}
\end{table}

\section{Quantitative Findings} 

\begin{figure}
\includegraphics[width=0.8\textwidth]{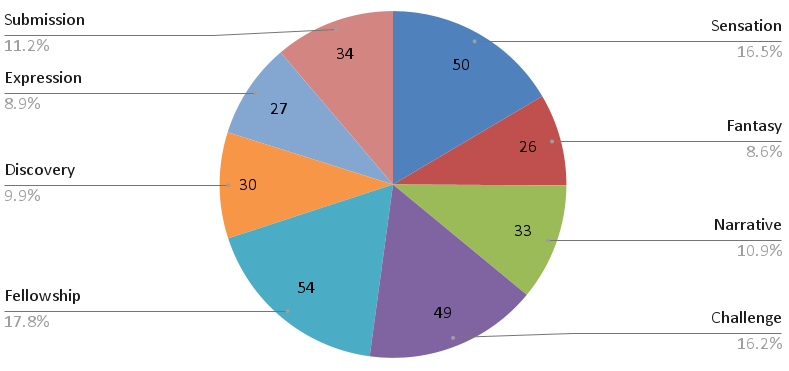}
\centering
\caption{Distribution of children's design ideas across 8 game pleasures. Numbers on the pie denote count of references. Percentages denote count relative to total references.} \label{8fun_pie}
\end{figure}

Each game design idea was coded to 1-7 game pleasures, yielding a total of 303 references. Among those, the three most prevalent categories of fun that emerged from our analysis were Fellowship, Sensation, and Challenge, each featured in over 16\% of the game ideas (see Fig. \ref{8fun_pie}). Nevertheless, all categories of game pleasures were represented, showing a balanced spread of fun across the dataset. No type of fun was completely absent, or overwhelmingly dominant. Finally, although there may have been some differences, all eight categories of fun were evident in ideas from various age groups.

\vspace{-0.7cm}
\section{Qualitative Observations: Pleasures of Robot Games} 
\label{qualitative}
\vspace{-0.3cm}
A total of 28 themes emerged from qualitative analysis. Below we present the elements of fun we observed in each category of LeBlanc's taxonomy.
Since this taxonomy was originally developed for video games, we draw comparisons throughout to illustrate how social robots' unique features can elicit enjoyment.
\vspace{-0.5cm}
\subsection{Sensation}
\begin{figure}
\centering
\includegraphics[width=\textwidth]{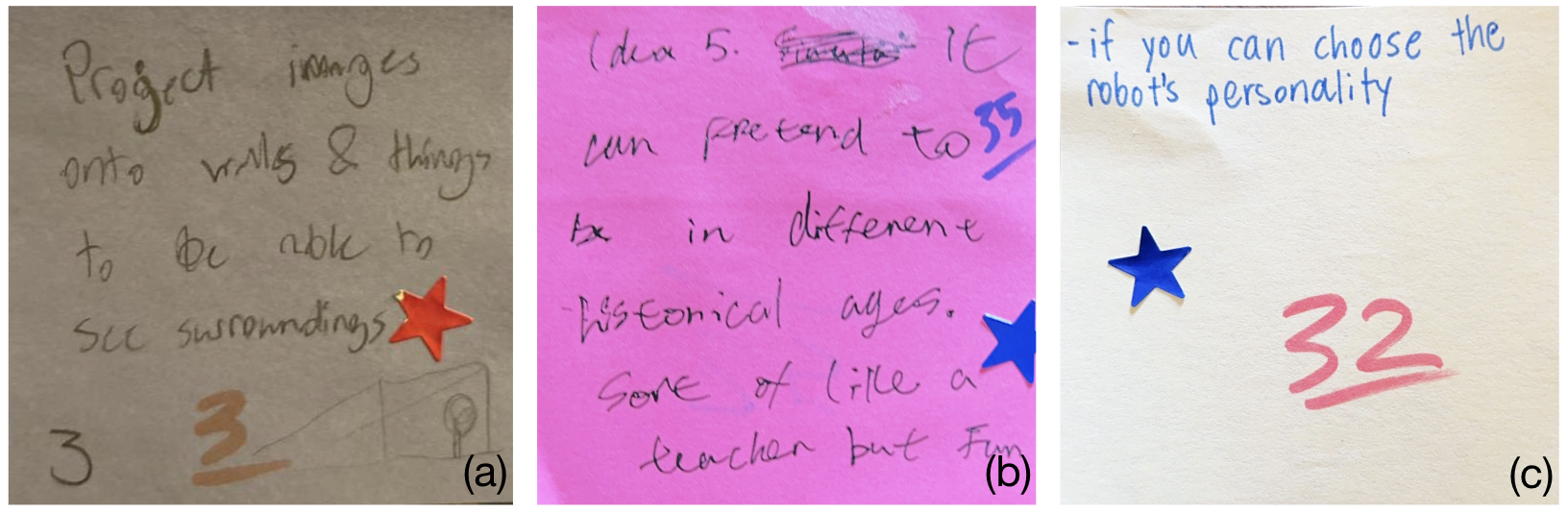}
\vspace{-0.7cm}
\caption{Sample ideas brainstormed by participants: (a):\textit{ ``Project images onto walls \& things to see surroundings.''}; (b): \textit{``Idea 5: it can pretend to be in different historical ages. Sort of like a teacher but fun''}; (c): \textit{`` if you can choose the robot's personality''}.} \label{fig:idea}
\end{figure}

\vspace{-0.1cm}
Compared to video games that are typically confined to 2D screens, robot games extend the game world into the player's 3D physical space, offering a more immersive gameplay. This allows for mixed reality interactions and use of physical props.
We identified 4 pleasures of robot games related to sensation:
\begin{itemize}
    \item \textbf{Dramatic Resonance}: Participants enjoyed the dramatic embodiment of fictional roles, and the ability to observe the robot embody different roles. 
    \item \textbf{Embodied Gameplay}: Many ideas incorporated physical activity (e.g., dancing), alluding to games that blend digital and physical play.
    \item \textbf{Use of Props}: Children frequently mentioned board games and card games, indicating a preference for tactile experiences and props.
    \item \textbf{World Aesthetics}: Some ideas emphasized the look and feel of the game world, enhancing the enjoyment of gameplay (see Fig. \ref{fig:idea}-a).
\end{itemize}
\vspace{-0.5cm}

\vspace{-1cm}
\subsection{Fantasy}
\vspace{-0.1cm}
Our observations indicate that robots can add unique fictional dimensions to games. Children perceived robots' non-human features as extraordinary superpowers and seemed to find the futuristic technological appeal of robots fascinating. We identified 4 types of fun related to fantasy:

\begin{itemize}
    \item \textbf{Simulating life}: Children demonstrated a desire for simulating aspects of real life, such as using the robot for ‘becoming’ different characters, or projecting their self onto the robot (e.g., \textit{``Robot acting out you in the past''}).
    \item \textbf{Hero's Journey}: Some ideas involved the robot as a hero with a mission (e.g., \textit{``Special gem holding together the world but people want to take it''}).
    \item \textbf{Non-Human Odyssey}: Participants were interested in exploring the robot’s non-human characteristics and affordances, e.g., changing its face.
    \item \textbf{Sci-Fi Companions}: Many ideas explored sci-fi-inspired fantasies of advanced robots serving as children's sidekicks and assistants.
\end{itemize}

\vspace{-0.5cm}
\subsection{Narrative}
Compared to video games, robots could introduce distinctive mediums of storytelling that children find enjoyable, including storytelling via their physical affordances (e.g., color of LEDs), as well as their social interactions (e.g., using gaze). We identified four types of fun related to narrative in robot games:

\begin{itemize}
    \item \textbf{Environmental Storytelling}: Children's ideas involved themes of robot interactions with the game environment to subtly convey narratives (e.g.,\textit{``Robot showing another robot how to watch clouds''}).
    \item \textbf{Robot-Mediated Storytelling}: Children enjoyed using unique robot affordances for storytelling; for example, by suggesting to change its accent based on scenarios, or change the color of its face when it gets angry.
    \item \textbf{Interactive Storytelling}: Children indicated a desire for having agency in shaping game narrative through making meaningful choices, dialogues with the robot, or generating their own stories from scratch.
\end{itemize}

\vspace{-0.5cm}
\subsection{Challenge}
Our observations indicated that children's perceptions of robots' social awareness often exceeded their actual capabilities. For instance, one child believed that robot could read their emotions because it has a camera. 
We argue that human-like features and social interactivity of robots can add a social twist to traditional challenges offered by games. We identified four kinds of challenge-related fun:
\begin{itemize}
    \item \textbf{Brainteasers}: A number of children's ideas involved playing guessing games such as puzzles, riddles, trivia, and quizzes with a robot (e.g., \textit{``guess who''}).
    \item \textbf{Reflex and Reaction}: Children indicated that robot games requiring quick thinking and reflexive responses would be fun 
    (e.g., \textit{``A game where Furhat winks or blinks and then you do the same thing, but if you do 1 wrong, you lose. Furhat does it very fast''} [sic]).
    \item \textbf{Mastery of Game Mechanics}: Children's ideas emphasized the importance of clear goals and feedback for player achievement, points, racing, and game levels (e.g., \textit{``Games are more fun if there are points''}).
    \item \textbf{Social Strategy}: Some ideas required using social skills to overcome obstacles, or help the robot (e.g., \textit{``Trick [another robot] into not bullying''}).
\end{itemize}

\vspace{-0.4cm}
\subsection{Fellowship}
This category of game pleasures is inherently social, making social robots an apt medium for enhancing fellowship aspect of games. Our findings show that fellowship was the most prevalent category of fun in children's game concepts. Children envisioned robots in various social roles, including co-player, avatar, and user interface for playing with other children. Using a robot as the player's embodied avatar is a unique feature of robot games that cannot be replicated with human co-players. You can embody a robot, but you cannot embody another person. This helps create aesthetic distance and encourages perspective-taking~\cite{sanoubari2022robot-mediated}. 
We identified four types of fun related to fellowship in robot games:
\begin{itemize}
    \item \textbf{Robot as Friend}: Some participants envisioned the robot as a co-player in games (e.g., acting as a \textit{``Simulated friend''}).
    \item \textbf{Robot as Avatar}: Some ideas involved using the robot as an avatar representation of the child in the game (e.g., \textit{``It can say words that I say''}).
    \item \textbf{Robot as Game}: In some ideas robots served as an interface facilitating gameplay (e.g., acting as a conversational interface in \textit{``Simon says'' \footnote{A children's game where players follow commands given by a leader, `Simon,' but only if the command begins with `Simon says,' such as `Simon says touch your nose.'})}.
    \item \textbf{Community}: Children indicated that interacting with other human players (e.g., via playing a game as a group) can elicit fun.
\end{itemize}

\vspace{-0.4cm}
\subsection{Discovery}
Unlike video games, robot can facilitate gameplay involving exploration of the physical world, as well as discovering the robot's capabilities.
It appears that the ambiguity of robots' capabilities sparks curiosity and drives children to playfully experiment to find robots' boundaries and fill in the gaps of their mental models. We identified three types of discovery fun that robot games could incorporate: 
\begin{itemize}
    \item \textbf{World Exploration}: Children expressed a desire to explore game worlds, both physically (e.g., \textit{``Hide and seek''}) and fictionally (see Fig.\ref{fig:idea}-b).
    \item \textbf{Game Discovery}: Some game concepts suggested that discovering new elements, levels,solutions, characters, or possible endings can elicit fun.
    \item \textbf{Robot Discovery}: Children found it fun to playfully experiment with a robot's capabilities and test its limits (e.g., limits of its face tracking system).
\end{itemize}

\vspace{-0.5cm}
\subsection{Expression}
Players value self-expression and enjoy making meaningful choices that have visible consequences \cite{schell2008art}. A distinct value that social robots can add to this category of fun is by acting as a vehicle for role-play \cite{sanoubari2022robot-mediated}.
We identified three types of fun related to expression that can be incorporated in robot games:
\begin{itemize}
    \item \textbf{Robot-Mediated Role-Play}: Many game ideas features role-playing with the robot or using the robots to assume different make-believe roles. 
    \item \textbf{Improvisation}: Several ideas involved improvisational games (e.g., \textit{``Charades with Furhat''}), suggesting that spontaneous creativity can elicit fun.
    \item \textbf{Robot Customization}: Participants indicated that personalizing the robot's appearance or characteristics can be enjoyable (e.g., see Fig.\ref{fig:idea}-c).
\end{itemize}

\vspace{-0.5cm}
\subsection{Submission}
Observations suggest that robots foster a suspension of disbelief, making gameplay more immersive. Three types of fun were identified in this category:
\begin{itemize}
    \item \textbf{Suspension of Disbelief}: Some ideas indicated that children enjoy temporarily setting aside knowledge or awareness of reality to embrace fictional elements (e.g., \textit{``simulating''} different people).
    \item \textbf{Robot Novelty}: Some children, particularly younger ones, seemed to find robot's novelty appealing, even in mundane tasks such as counting cards.
    \item \textbf{Robot Control}: Children seemed to find exerting control over a robot's behavior inherently enjoyable and exciting.
\end{itemize}

\section{Informing the Design of \REMind}
Informed by these observations, we refined the design of \REMind{} game mechanics and narrative to incorporate some of the identified elements of fun.
A high level summary of the updated game narrative is as follows. 

Upon entering the game world, the player learns that a pervasive \textit{glitch} has affected the Furhat robots in the fictional world of \textit{Kernel}. This glitch, caused by an unknown algorithm the robots have learned, is increasingly dangerous and detrimental. It causes the robots of Kernel to exhibit signs of disharmony and confusion. For example, they forget small things, such as parts their sentences or answers to simple questions, their speech becomes jumbled and incoherent, and their projected faces show visual distortions.

To eliminate the glitch and restore balance in Kernel, the main robot character, ECHO, must embark on a heroic journey to its own past, simulate the memory that started the glitch, and overwrite it. However, ECHO can only achieve this with help from a human child, because solving the glitch requires one to reason using their emotions-- something robots lack. ECHO and the player discover that the critical memory involves a peer bullying incident, enacted by FUSE and JITTER, two other Furhat robots in the room. In this memory ECHO was a bystander that observed FUSE bully JITTER and did not take any action.

\begin{figure}
\includegraphics[width=0.7\textwidth]{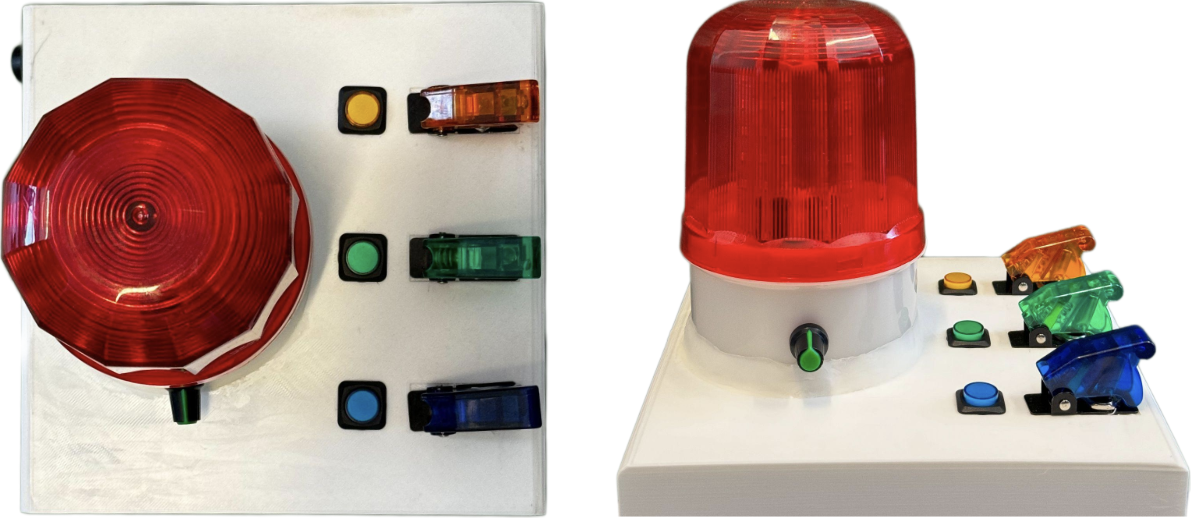}
\centering
\vspace{-0.2cm}
\caption{``Bullying Detection Circuit'': A prop to teach children about the three criteria for bullying. The pushbuttons trigger recorded messages explaining each criterion.} \label{Circuit}
\end{figure}

To overwrite the memory, the player engages in a series of problem-solving steps. First, the player helps ECHO take the perspectives of FUSE and JITTER by reflecting on their emotions and intentions. Then, the player helps ECHO reflect on why it did not intervene. ECHO reveals its ``Bullying Detection Circuit'' (see Fig. \ref{Circuit}) did not sound an alarm, leading it to believe that the incident was merely teasing. The player searches the room to find the circuit and test it. By exploring this prop, the player discovers that when all three switches (representing the criteria for bullying outlined in section \ref{intro_to_REMind}) are turned on, the beacon light on the circuit flashes, indicating bullying has been "detected". The player also finds out that the alarm has been muted and uses the adjustable volume knob to fix the circuit, preventing future bullying from going unnoticed.
Next, the player helps ECHO overwrite the glitchy memory by choosing one of the safe intervention strategies outlined in section \ref{intro_to_REMind} and showing the robot how to do it. This involves the player improvising what ECHO should say, and demonstrating it to the robot. The player is guided to record themselves enacting what the intervention. Their facial expressions and voice are then streamed onto the Furhat robot so the player can watch their enactment embodied by the robot ECHO and observe how the story unfolds as a result.

This refined narrative aligns \REMind{} game mechanics to learning goals outlined in section \ref{intro_to_REMind} and incorporates elements of fun. Table \ref{tab:fun_elements_in_REMind} provides examples of how specific elements of robot game pleasures were implemented in \REMind{}.

\begin{table}[h!]
\centering
\caption{Robot game pleasures and examples of how they informed \REMind{} design}\label{tab:fun_elements_in_REMind}
\vspace{-0.2cm}
\resizebox{1\columnwidth}{!}{%
\begin{tabular}{lll}
\hline
\textbf{}  & \textbf{Robot Game Pleasures} & \textbf{Example Implementation in REMind}        \\ \hline
Sensation  & Dramatic Resonance            & - Interactive role-play via Furhat robots              \\
           & Embodied Gameplay             & - Player explores the room and interacts with props                 \\
           & Use of Props                  & - Use of ``bullying detection circuit''            \\
           & World Aesthetics              & - Use of room-based light and audio cues             \\ \hline
Fantasy    & Simulating life               & - Use of robots to explore peer bullying      \\
           & Hero's Journey                & - ECHO's journey to eliminate the glitch             \\
           & Non-Human Odyssey             & - ECHO overwriting a past memory          \\
           & Sci-Fi Companions             & - Player explores a fictional robot universe         \\ \hline
Narrative  & Environmental Storytelling    & - Light and sound cues aiding storytelling       \\
           & Robot-Mediated Storytelling   & - Glitch distorts robots' speech and face       \\
           & Interactive Storytelling      & - Player decides how ECHO intervenes         \\ \hline
Challenge  & Brainteasers                  & - Player troubleshoots the glitch        \\
           & Reflex and Reaction           & - Use of time limits for challenges                         \\
           & Mastery of Game Mechanics     & - Use of progressive levels for resolving the glitch \\
           & Social Strategy               & - Player strategizes against peer bullying    \\ \hline
Fellowship & Robot as Friend               & - Player and robot collaborate to fix the glitch \\
           & Robot as Avatar               & - ECHO is used as player's avatar        \\
           & Robot as Game                 & - Two robots are used as non-player characters (NPC)                 \\
           & Community                     & - All players meet for post-game debrief      \\ \hline
Discovery  & World Exploration             & - Player searches for objects in the room            \\
           & Game Discovery                & - Player discovers new challenges                 \\
           & Robot Discovery               & - Player discovers Furhat capabilities         \\ \hline
Expression & Robot-Mediated Role-Play      & - Player role-plays by using ECHO as a proxy                 \\
           & Improvisation                 & - Player determines how ECHO intervenes      \\
           & Robot Customization           & - Player customizes ECHO as avatar          \\ \hline
Submission & Suspension of Disbelief       & - Player engages with a fictional story       \\
           & Robot Novelty                 & - Player interacts with Furhat robots         \\
           & Robot Control                 & - Player's recorded gesture is streamed onto ECHO      \\ \hline
\end{tabular}%
}
\vspace{-0.5cm}
\end{table}

\section{Discussion} 
\vspace{-0.3cm}
We discussed that to be effective, educational games must be fun for their audience. We aimed to understand how to design an anti-bullying robot game called \REMind{} such that it is fun for children, while addressing its learning goals. We conducted brainstorming sessions with focus groups of children and analyzed their ideas using Framework Analysis~\cite{goldsmith2021using}, leveraging LeBlanc's Taxonomy of Game Pleasures~\cite{hunicke2004mda}. Our analysis revealed 28 elements that can be leveraged to create fun experiences with social robots, which informed the design of \REMind. We offer three recommendations for designing social robot GBL experiences.

First, we argue that designers should move beyond creating robot \textit{interactions} and instead use robots to create holistic mixed reality~\cite{speicher2019mixed} \textit{experiences}. Games provide a systemic framework and invaluable tools for this purpose. Notably, Elemental Tetrad of Games~\cite{schell2008art}, comprises of four interconnected part: mechanics, aesthetics, story, and technology. Mechanics define the game's rules and actions; aesthetics encompass its visual, auditory, and sensory appeal; story provides narrative context; and technology enables the game's creation and functionality. These elements work in harmony to shape the overall player experience. For example, \REMind{} uses Robot-Mediated Storytelling to convey robot speech getting jumbled as a symptom of the glitch, which is achieved by using a voice transformation function to increasingly replace ECHO's speech stream with a set of nonsensical words during glitches. This blends the mechanics of worsening glitches, the aesthetic of distorted speech, the narrative of glitch-affected robots, and the technology of voice modulation. We recommend that designers focus on leveraging stories and bringing them to life by using social robots.

Second, facilitating playful learning depends on a break from reality. The `magic circle' in games refers to the space, both physical and mental, where the fictional rules and reality of the game world are temporarily accepted, allowing participants to experiment freely without fear of failure. Piaget and Vygotsky have both emphasized the role of play in allowing children to step away from the \textit{``here and now,''} fostering deeper learning \cite{plass2020handbook}. Creating this magic circle requires little technology as fiction is built on agreed-upon rules and constraints. It is these rules that allow a stick to become a sword and a blanket to become a fortress in children's games of make-believe. Designers of robot transformational games should leverage the power of make-believe. 

Finally, we recommend that designers prioritize interactivity in the game, as fun and interactivity often go hand in hand. Game designers use various interactive loops, like action-feedback, cognitive, emotional, and social interactivity, to create engagement \cite{sellers2017advanced}. In this study, we observed that children do not necessarily need the robot to perform extraordinary actions to have fun. Instead, many of their ideas for making the game enjoyable involved the robot prompting them to take actions (e.g., by playing \textit{``Simon says''}). Therefore, designers should focus on creating interactive experiences with clear goals, varied activities, and immediate feedback, rather than having children passively observe robot's actions.

This study's limitations include the small sample size, limited discussion time for children's ideas, and the wide age range of participants. Additionally, the brainstormed ideas, tailored to Furhat robots' unique features, may not be generalizable to other robotic platforms. Methodological constraints, such as not tracking participant IDs and limited discussion of age and group differences, further impact the clarity and generalizability of the results. Future work will focus on implementing \REMind{} and evaluating its effectiveness through play-testing sessions with children.

\vspace{-0.2cm}
\section{Conclusion}
This paper explores how to design fun, game-based educational experiences using social robots. We conducted participatory design research with children to understand what they find fun in a transformational robot game, categorized their responses using LeBlanc's Taxonomy of Game Pleasures, and identified 28 elements of fun which informed the design of an anti-bullying game called \REMind.
We argue that social robots are uniquely positioned to enhance learning through play because of their ability for embodied social interaction. This research aims to help designers create engaging and effective transformational games with robots.
%
\begin{credits}
\subsubsection{\ackname} 
We would like to thank Dr. Christina Salmivalli and Dr. Claire Garandeau, psychologists with expertise in peer relations and school bullying, for their valuable consultations that significantly contributed to \REMind{}.
This research was undertaken, in part, thanks to funding from the Canada 150 Research Chairs Program.
\end{credits}
\bibliographystyle{splncs04}
\bibliography{refs}
\end{document}